\documentclass[aps,singlespace,showpacs,preprintnumbers,amsmath,amssymb,letterpaper,prb]{revtex4}

\usepackage{graphicx} 
\usepackage{dcolumn} 
\usepackage{bm} 
\usepackage{amsmath}

\begin{document}


\title{Polarization and Charge Transfer in the Hydration \\ of Chloride Ions}

\author{Zhen Zhao and David M. Rogers\footnote{Present address: Sandia National Laboratories, MS 1314, PO Box 5800, Albuquerque, NM 87185}}
\affiliation{Department of Chemistry, University of Cincinnati, \\
Cincinnati, OH 45221-0172}
\author{Thomas L. Beck}
\email{thomas.beck@uc.edu}
\affiliation{Departments of Chemistry and Physics, University of Cincinnati, \\
Cincinnati, OH 45221-0172}

\date{\today}

\begin{abstract}
A theoretical study of the structural and electronic
properties of the chloride ion and water molecules in the first hydration shell is presented. 
The calculations are performed on an ensemble of configurations obtained from molecular
dynamics simulations of a single chloride ion in bulk water. The simulations utilize the polarizable 
AMOEBA force field for trajectory generation, 
and MP2-level calculations are performed to examine the electronic structure properties of the ions 
and surrounding waters in the external field of more distant waters. The ChelpG
 method is employed to explore the effective charges and 
dipoles on the chloride ions and first-shell waters. The Quantum Theory of Atoms in Molecules (QTAIM) is further
utilized to examine charge transfer from the anion to surrounding water molecules.
The clusters extracted from the AMOEBA simulations exhibit high probabilities
of anisotropic solvation for chloride ions in bulk water. 
 From the QTAIM analysis, 0.2 elementary charges are transferred
from the ion to the first-shell water molecules.  The default AMOEBA model
overestimates the average dipole moment magnitude of the ion compared with the
estimated quantum mechanical value.   The average
magnitude of the dipole moment of the water molecules in the first
shell treated at the MP2 level,
with the more distant waters handled with an AMOEBA effective charge model, is 2.67 D. This value
is close to the AMOEBA result for first-shell waters (2.72 D) and is slightly reduced from the bulk AMOEBA value (2.78 D). 
The magnitude of the dipole moment of the water molecules in
the first solvation shell is most strongly affected by the local water-water
interactions and hydrogen bonds with the second solvation shell, rather than by interactions with the ion. 
\end{abstract}
\pacs{82.60.Lf,87.16.A-,61.20.Ja,64.70.qd,64.75.Bc}
\keywords{Quasi-chemical Theory, Hydration}

\maketitle

\section{Introduction}{\label{sec_Intro}}

Fundamental studies of the thermodynamic and structural properties of ions
in water and near proteins are important for understanding a wide range of chemical and 
biological phenomena. For example, in the central
nervous system, ion-coupled membrane transporters regulate signaling
by utilizing the electrochemical potential of specific ions to pump
organic substrates and amino acids across the cell
membrane.\cite{stein,hille,defelice} As another example, chloride transporters exchange two chloride ions for one proton during the transport cycle.\cite{accardi_secondary_2004,yin_ion_2004,kuang_proton_2007} To gain more insight into the mechanism, specificity, and function
of these transporters, the binding properties of specific ions to the transporters and the ion hydration process in bulk water need to be characterized. 

The hydration of atomic and molecular ions has been the focus of intensive research for over 100 years. More recently, specific ion effects have resurfaced in diverse fields,\cite{wkunz04} including hydration free energies, ion activities, surface tension increments, bubble interactions, colloid interactions, biological membrane multilayer swelling,\cite{petrache_salt_2006} and 
polymer phase equilibria,\cite{zhang_interactions_2006} to name several examples. Such specificity requires theoretical treatments that go beyond simple dielectric models.  Specific interactions at the molecular level are involved, and that specificity can lead to dramatic changes in bulk properties when one ion is substituted with another.\cite{petrache_salt_2006} It has been argued that ion specificity is a central problem in connecting physical science to biological systems, and that the connection has not been fully made so far.\cite{ninham_building_2005}   

Calculation of the solvation and binding properties of an ion in water
and near proteins for a classical model is now routine on modern
workstations. The accuracy of these
calculations depends sensitively on the classical model potential used, however. 
In the same spirit as Doren, Wood, and coworkers,\cite{wood1,wood2,wood3,wood4} we are
interested in incorporating quantum mechanical calculations in the
study of the thermodynamics of ions in water and near proteins. The basic idea in Refs.~\cite{wood1,wood2,wood3,wood4} is to perform classical simulations, obtain the free energy from
those classical simulations, and then use the classical configurations coupled with statistical mechanical perturbation theory to
correct those free energies towards the quantum result. This method is less
expensive than a direct {\it ab initio} molecular dynamics (AIMD) quantum simulation\cite{laasonen,heuft_density_2003,eguar09,whitfield_theoretical_2007}  since it uses a classical
simulation to generate configurations for thermal averages. It is also more
accurate than a classical simulation since it can in principle represent the electronic
properties near the ions from first principles. 

As a first step, here we study the aqueous solvation structures of the chloride ion in
bulk water by utilizing the classical polarizable AMOEBA force field simulations and then by performing
detailed quantum mechanical calculations on local solvation clusters
extracted from the classical simulations. The cluster refers to the
chloride ion and the coordinating water molecules in the first
solvation shell, including interactions with more distant waters at the AMOEBA level. 
The polarization of the water molecules and ions in their
solvation environment is explored by analyzing charges and dipoles generated from the charge distribution in the
quantum mechanical model. Higher{}-level electronic structure methods (MP2 level) are employed to include electron correlation effects at a modest level of accuracy. Also, the first-shell water polarization is studied as a function of the cluster size. The present
approach allows efficient calculation of configurational averages for an accurate
QM/MM model. This
paper is part of a series developing and exploiting computational
methods for calculating the electronic and thermodynamic
properties of ions in water.\cite{droge08,droge09}  Future work will focus on ion binding in proteins. 

The paper is organized as follows.  We next discuss the classical and quantum computational approaches 
employed for the study of local hydration structure.  The results of the calculations are then presented, followed
by our conclusions and discussion of implications for simulations of ion hydration and future research directions. 

\section{theoretical and computational methods}
\label{sec_Methods}

In this section, we present our approach for studying the electronic
structural properties of the anion solvation shell.

\subsection{Classical Simulation of the Chloride Ion in Water }

The molecular dynamics (MD) trajectories were generated using the polarizable AMOEBA force
field\cite{ren2,ren3,ren4}  as implemented in the AMBER package.\cite{amber10} The
AMOEBA force field has been shown to reproduce the expected dipole
moment, dielectric constant, and energetics of bulk water from the gas phase
to the bulk phase\cite{ren3} and over a wide range of temperatures and pressures.\cite{ren2}
In addition, this force field has been extended to include ion-water interactions, and 
excellent results for ion hydration free energies have been obtained.\cite{grossfield2}

The simulated bulk system consists of a single Cl$^-$ ion and
215 water molecules. The simulations were performed using a timestep of
1.25 fs. Configurations were saved every 0.5 ps for later analysis. 
Periodic boundary conditions were applied in all three dimensions. Long-range 
electrostatic interactions were handled using the PME algorithm\cite{essmann} with a
real{}-space cutoff of 8 {\AA}. The nonbonded interactions were
truncated at 10 {\AA}. The temperature was maintained at 300 K using
Langevin dynamics with a collision frequency of
5ps\textsuperscript{{}-1}. \ The pressure was maintained at 1.0 atm
using the isotropic position scaling scheme with a pressure relaxation
time of 2.0 ps. The cubic simulation box was allowed to scale in size
isotropically in order to maintain constant pressure. The system was
equilibrated for 120 ps (the system density was seen to converge within
the first 20 ps), and then a production run of 500ps was performed for
further analysis.

\subsection{Electronic Structure Calculations}

The geometries of the anion-water clusters were extracted from the AMOEBA molecular
dynamics trajectories for the subsequent electronic structure calculations.  
The clusters for a given system configuration consisted of the chloride ion and all waters that satisfied 
a hydrogen-bonding condition discussed below.
There is of course no guarantee that these configurations
accurately represent the solvation environment at the quantum level.   In another paper,\cite{droge09} 
we have found that anion solvation anisotropy in classical polarizable force field simulations is significantly 
affected by the polarizability of the anion.  In addition, we found that the AMOEBA model over-polarizes the chloride ion 
by roughly a factor of 2 relative to AIMD simulations.\cite{eguar09,droge09}  Thus we performed simulations both with the default 
AMOEBA chloride ion polarizability (4 {\AA}$^3$) and 
with the polarizability reduced by a factor of 2.  While the solvation anisotropy decreases with the lower polarizability value, the results for
water dipole magnitudes and charge transfer computed quantum mechanically did not substantially change; as expected, the ion dipoles 
from the AMOEBA model decreased substantially to 62\% of the computed average from the default simulation, and the estimated quantum mechanical ion dipoles
decreased slightly due to the reduced hydration anisotropy (below).
  
The electronic structure calculations were performed at the 2nd-order perturbation theory
level.  The MP2 calculations were performed with the Gaussian 03 package\cite{g03}
using augmented correlation{}-consistent polarized valence basis
sets of double ${\zeta}$ quality (aug{}-cc{}-pVDZ).\cite{dun4} This basis set has been shown to yield accurate molecular electrostatic interaction energies\cite{volkov} and charge transfer estimates 
compared with experiment.\cite{szefczyk} The total energy was
converged to a precision of 10\textsuperscript{{}-8} Hartrees. The
charge distribution was investigated using the same method and basis
set. 

\begin{table}
\caption{The effect of basis set on the dipole moment of an
isolated water molecule 
calculated at the MP2 level.}
\begin{tabular}{cc}
Basis Set & ${\mu}_{SCF}$(Debyes) \\ 
\hline
6-31G & 2.54220  \\
6-31G** & 2.09900  \\
6-31+G* & 2.33330  \\
6-31+G** & 2.24850  \\
6-31++G** & 2.23730  \\
6-31++G(2df,p) & 2.03720  \\
6-31++G(3df,2p) & 1.88510  \\
6-311++G(3df,2p) & 1.93650  \\
cc-pvtz & 1.92060  \\
aug-cc-pvdz & 1.86620 \\
aug-cc-pvtz & 1.85120  \\
\hline
\end{tabular}
\label{tab:basis}
\end{table}

Table \ref{tab:basis} displays the convergence of the MP2-level single-water 
dipole moments with increasingly accurate basis sets. The experimental geometry for the monomer [r[OH]=0.9572,
{\textless}HOH = 104.52] was assumed.\cite{benedict}  The dipole moment 
of the monomer calculated with Dunning's
correlation consistent basis sets augmented with diffuse functions
(aug{}-cc{}-pvdz and aug{}-cc{}-pvtz) shows the best agreement with the
experimental result (1.85-1.86 D).\cite{lovas} We chose the aug{}-cc{}-pvdz basis
set\cite{dun4} for both 
efficiency and high quality results.\cite{volkov,szefczyk}

 There exists a variety of methods available for estimating 
atomic charges and dipoles from the electron density. 
One approach is to obtain effective charges from the Electrostatic
Potential (ESP) by fitting charges (and perhaps point dipoles) to match the ESP 
computed at the quantum level; points within the van der Waals radii of 
the atoms in the molecule or cluster are excluded from the fit.  The ChelpG method is a widely
used numerical implementation of this approach.\cite{breneman} 
The ESP methods have been applied in 
Monte Carlo simulations,\cite{nagy} molecular dynamics simulations,\cite{mayaan,holm_09}
 molecular modeling,\cite{kamath} and other applications.\cite{katritzky} Besides the
ESP methods, atomic charges can be determined more rigorously by
Bader's Quantum Theory of Atoms in Molecules
(QTAIM).\cite{bader1,bader2}  The QTAIM approach provides an unambiguous quantum definition of an atom within 
a molecule based on the zero-flux surface of the electron density surrounding 
the atom (or ion). The zero-flux condition in addition leads to the result that the total energy of the system can be expressed as the sum of the atomic energies.  Due to its fundamental quantum mechanical underpinnings, the QTAIM method would appear to provide the least ambiguous definition of atomic charges in molecules and/or clusters.

Studies have shown that the ChelpG method is capable of providing accurate estimates
of molecular dipoles\cite{maciel,gomes,martin_04} and charge transfer effects.\cite{szefczyk} The results
we present below show that the ChelpG-estimated anion and first-shell water dipoles are physically 
consistent with other quantum methods such as AIMD simulations.  The estimated net effective charge on 
the chloride ion is somewhat ambiguous from the ChelpG calculations, however, since a single charge is 
employed in the ESP fitting to represent the complicated and diffuse charge distribution of the anion.  
Concerns have also been raised related to applying 
the ChelpG method to densely packed systems involving charged interactions occurring
within the van der Waals radii of the interacting species;\cite{masamura_00,gomes} recent
work has shown, however, that accurate estimates of effective atomic charges can be obtained 
for dense ionic systems
in comparison with an alternative robust charge placement (Bl\"ochl) algorithm.\cite{holm_09,blochl} 
Our ChelpG cluster studies involve the chloride ion and up to 6 water molecules, and the ion is most often
near the surface of the cluster. Thus there are not many `buried' atoms in the ESP fitting, a point
of concern in previous studies.\cite{holm_09}   
In light of these ambiguities, we further 
analyzed possible charge transfer effects with the QTAIM method by computing the distribution of instantaneous anion charges 
within the surface centered on the chloride ion and specified by the zero-flux condition $\nabla \rho \cdot {\bf n} = 0 $.   
Both the ChelpG and QTAIM methods appear to be relatively
insensitive to basis set errors,\cite{tsuzuki_96,szefczyk,martin_04} unlike the Mulliken population
analysis.  

The ChelpG atomic charges were calculated
from the electronic density using the ChelpG routine of the Gaussian
package, imposing the restriction that the ChelpG total cluster dipole
moment reproduce the MP2/aug{}-cc{}-pvdz value
calculated directly from the electronic density.  For a discrete charge distribution
obtained from the ChelpG optimization, the dipole moment of a
molecule can be calculated in the standard way as a sum over the products of the effective atomic charges and
the atomic locations relative to a reference point; the reference point does not affect the dipole for a neutral molecule.
The ChelpG and QTAIM charge analyses below suggest, however, that there
is a certain degree of charge transfer between the ion and neighboring
water molecules. For example, on average, there is a charge transfer in
the amount of roughly 0.03-0.05e in a Cl$^-$/(H$_2$O)$_6$ cluster from the
Cl$^-$ ion to each coordinating water molecule. The dipole moment of a water molecule 
is not uniquely defined if the water
molecule is not neutral. Since the amount of charge transfer to each
water molecule is relatively small, we chose the water reference point to be
the oxygen atom of the considered water molecule.  
To obtain an estimate of the dipole magnitude of the anion, the ChelpG fit was extended to 
include a point dipole on the anion; all other ChelpG charge fits were performed with a single
charge on the chloride ion.
The QTAIM net ion charges
were computed with a code developed by Henkelman, Arnaldsson, and J\"onsson.\cite{henkelman} 

We utilized the AMOEBA multipole and induced dipole model for waters
beyond the first shell and implemented a localized effective charge
distribution to mimic the field generated by those fixed and induced multipoles; 
discrete charges were distributed near the location of the 
point dipoles and quadrupoles so as to accurately mimic the electrostatic potential away from the multipoles.  These charges
were then used to generate an external potential in the Gaussian code. We tested this model
against higher level \textit{ab} \textit{initio} calculations. Table \ref{tab:chgmodel}
shows the effect of the charge model on the ChelpG calculated dipole moment of
a water molecule in a water dimer, in which one water molecule is
treated on the MP2 level and the other one is represented by a charge
model. We found that all three water charge models (TIP3P, TIP4P, and AMOEBA)
provide a reasonable field to reproduce the dipole moment calculated at
the MP2 level. To be consistent with the molecular dynamics model, we chose the AMOEBA
model for the external field. 

\begin{table}
\caption{The effect of the charge model on the calculated dipole moment of a
 water molecule in a water dimer by the ChelpG charges, in which one water molecule is 
treated at the MP2 level and the other one is represented by a charge model. The final entry is the
result when both water molecules are treated at the MP2 level followed by ChelpG determination 
of the dipole of one of the waters.}
\begin{tabular}{cc}
Charge Model & $\mu_{ChelpG}$(Debyes)\\ 
\hline
TIP3P & 2.08 \\
TIP4P & 2.08 \\
AMOEBA & 2.09 \\
MP2 & 2.12 \\
\hline
\end{tabular}
\label{tab:chgmodel}
\end{table}

\section{Computational Results}
\label{sec_Results}

\subsection{Local hydration structure in AMOEBA simulations}

\begin{figure}
\includegraphics[angle=0,scale=1.4]{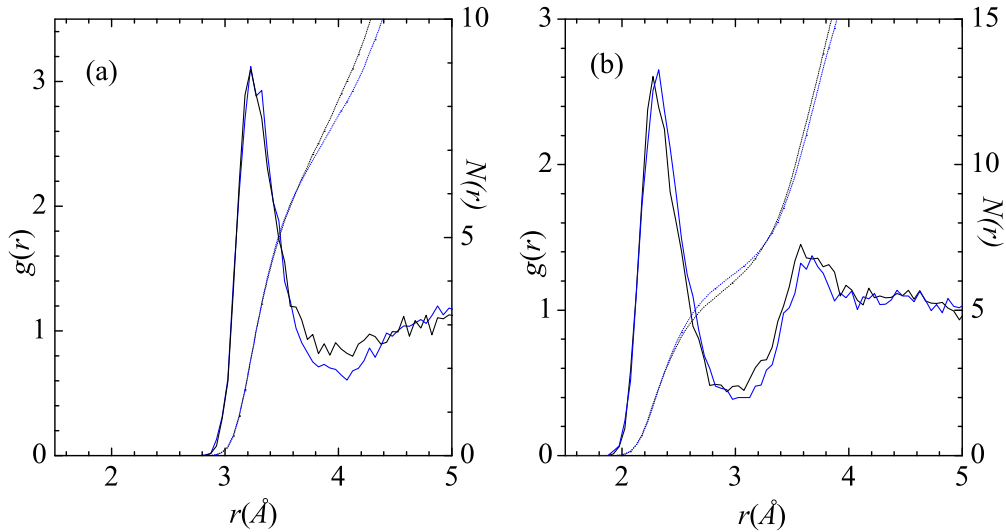}
\caption[Radial distribution] {Radial distribution functions $g(r)$ and integrated values $N(r)$ from the AMOEBA simulations: 
(a) chloride/(water oxygen) and (b) chloride/(water hydrogen). The black curves are for the default
chloride ion polarizability simulation, while the blue curves are for the reduced ion polarizability simulation.}
\label{fig:rdfs}
\end{figure}

\begin{figure}
\includegraphics[angle=0,scale=1.0]{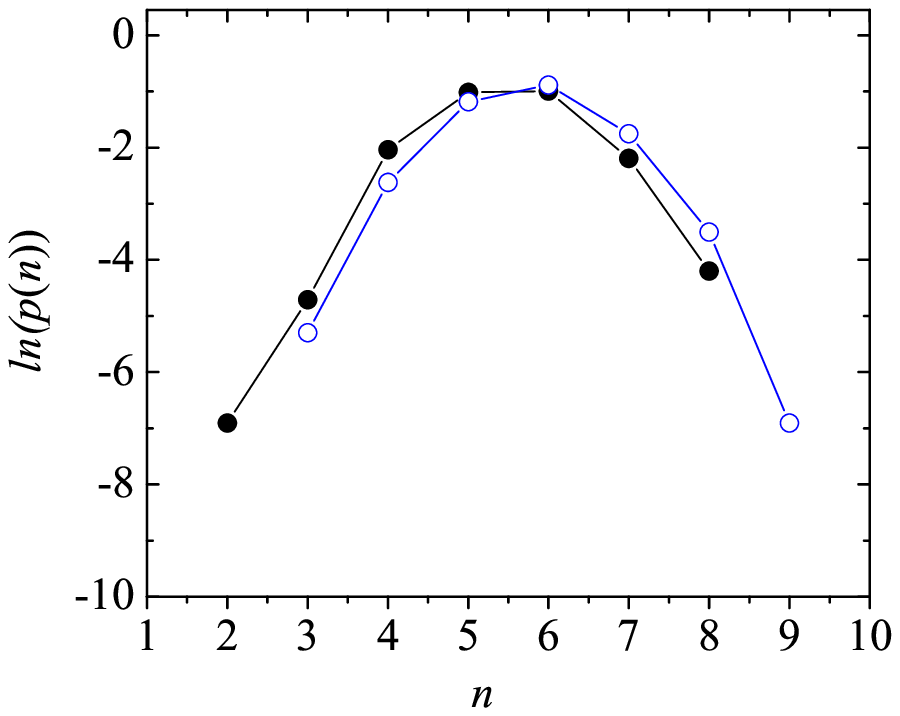}
\caption[Coordination distribution] {The log of the distribution of the coordination number \textit{n} of Cl
$^-$ (aq) from the AMOEBA simulation. The curve shown in black is for the default ion polarizability
simulation, while the blue curve is for the reduced ion polarizability simulation. } 
\label{fig:coord}
\end{figure}

The radial distribution functions (RDFs: \textit{g}(\textit{r})) between
the Cl$^-$ ion and water molecules (O and H atoms)
are shown in Figure \ref{fig:rdfs} for chloride ion polarizabilities of 4 (default) and 2 {\AA}$^3$.  The effect of 
the anion polarizability on the first peaks of the Cl$^-$-H and Cl$^-$-O RDFs is small; the first 
minimum in the Cl$^-$-O RDF for the 2 {\AA}$^3$ polarizability case becomes slightly deeper, however,
relative to the default polarizability value, likely due to a somewhat stronger attraction between the 
more-polarized ion and second-shell waters for the 4 {\AA}$^3$ polarizability case. 
The RDF for the Cl$^-$ ion and
oxygens of water shows a first peak at $\sim$3.2 {\AA}, which agrees
with the experimental Cl$^-${}-H$_2$O bond
length obtained by X{}-ray and neutron diffraction measurements.\cite{xray,neut} The
first peak has a distribution range from 2.9 {\AA} to $\sim 4.0$ {\AA}. The
RDF for the ion and hydrogens of water contains a nearest{}-neighbor
peak near 2.3 {\AA}. The first minimum of the Cl$^-$-H RDF occurs around
3 {\AA}. The coordination number for Cl$^-$ in H$_2$O
obtained by integration of the Cl$^-${}-H pair distribution
function up to the first minimum is 5.9 for the default chloride ion polarizability and 6.2 for the 
reduced polarizbility case. The instantaneous coordination
number is defined as the number of hydrogens within 3 {\AA} (the first
minimum in the Cl{}-H pair correlation function) with an
O{}-H{\textperiodcentered}{\textperiodcentered}{\textperiodcentered}Cl
angle greater than 130 $^{\circ}$ (a commonly used lower limit in
hydrogen bond analysis). Figure \ref{fig:coord} displays the log of the coordination number as a functinon of $n$ and shows that the instantaneous coordination number fluctuates in the range 2 to 9 with a maximum probability at 6. 
The average coordination number is 5.5 for the 4 {\AA}$^3$ polarizability case, slightly less than that obtained by
integration of the RDFs due to the angular hydrogen-bonding restriction.  Reduced anion polarizability shifts the coordination number
distribution to slightly larger coordination numbers, with an average value of 5.8. 
These structural results are qualitatively similar to
Car-Parrinello molecular dynamics density functional theory
(CPMD-DFT) simulations of the chloride and bromide ions in water.\cite{heuft_density_2003,raugei7} We note that coordination number disributions, along with occupancy numbers for waters in the same observation volume, can yield free energy differences between the various coordination states.\cite{lrp_coordination_09}
 
\begin{figure}
\includegraphics[angle=0,scale=.7]{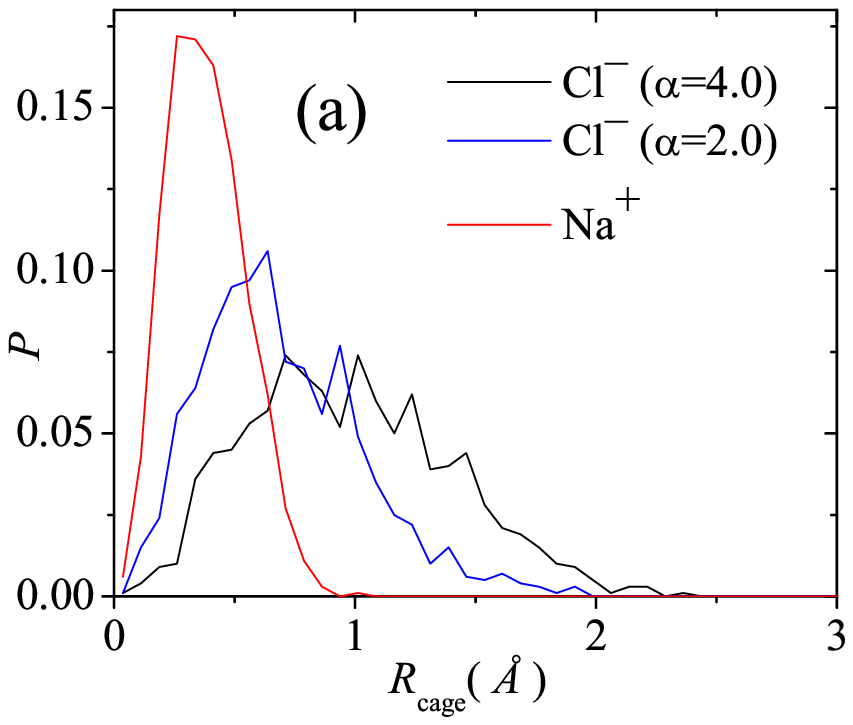} 
\includegraphics[angle=0,scale=.7]{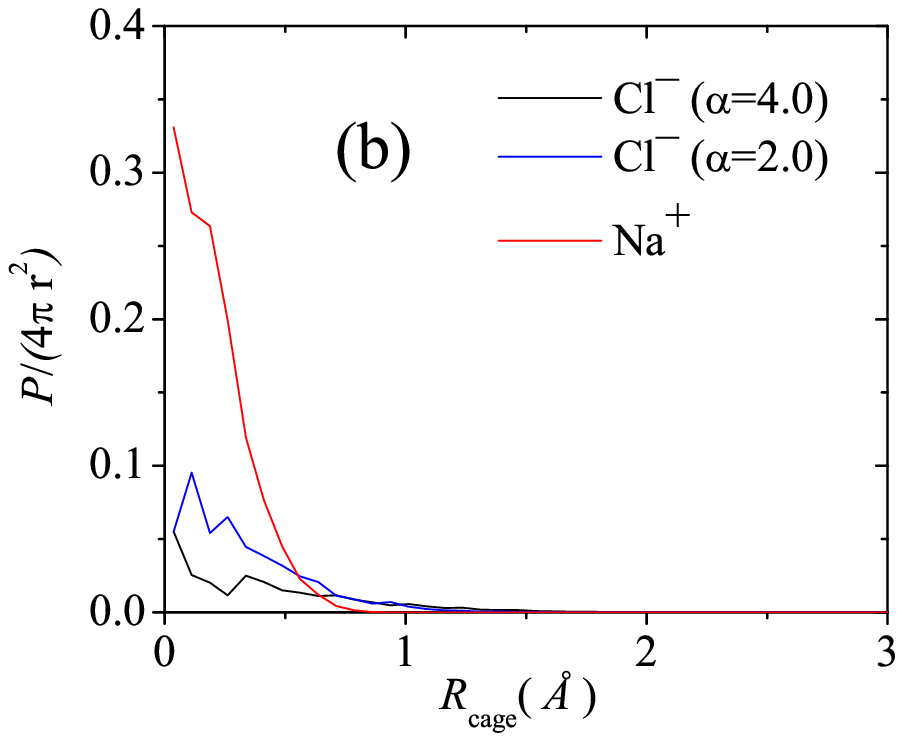}
\caption{(a) The distributions of $R_{\mathrm{cage}}$, the distance between the center
of mass of the first solvation shell water O atoms and the Na$^+$ and Cl$^-$ ions, from the AMOEBA simulations. (b) The
distributions divided by $4\pi R_{\mathrm{cage}}^2$. } 
\label{fig:cage}
\end{figure}

\begin{figure}
\includegraphics[angle=0,scale=0.8]{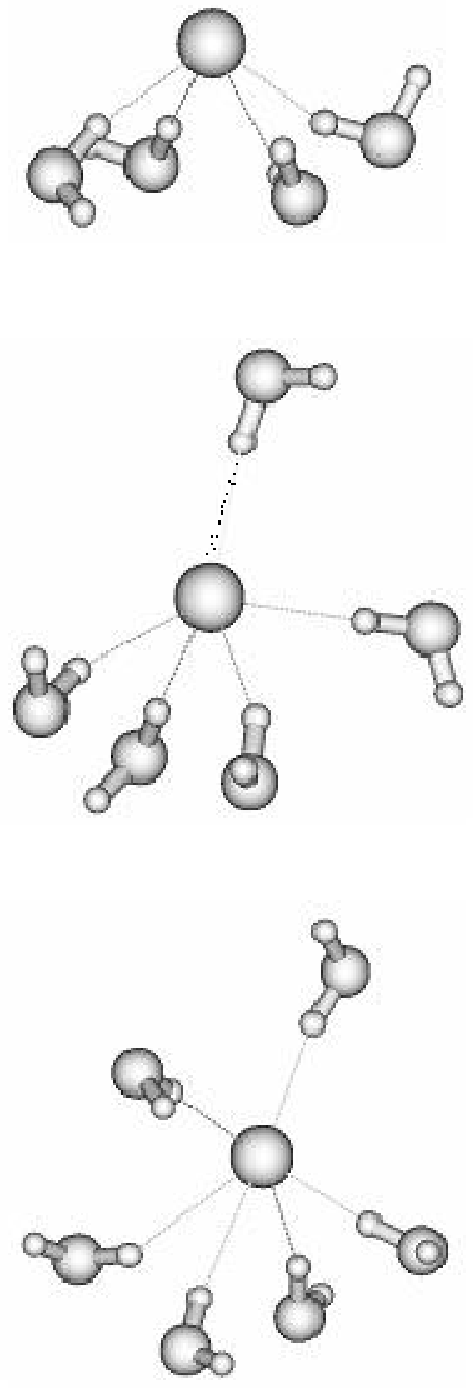}
\caption{Three snapshots of typical configurations of the Cl$^-$/H$_2$O first hydration shell
from a simulation with the AMOEBA force field (default ion polarizability).}
\label{fig:snap} 
\end{figure}

Figure \ref{fig:cage} shows that, for the default chloride ion polarizability simulation,
the distance between the center of
mass of first shell waters and the Cl$^-$ ion,
$R_{\mathrm{cage}}$, fluctuates in the range 0.1 to 2.5
{\AA}  with a
peak around 1 {\AA}. Comparison of the distribution with that for the Na$^+$ ion
suggests an increased anisotropy of the solvation
shell around the chloride ion. That anisotropy is reduced when the chloride ion polarizability is reduced by
a factor of two, consistent with our previous simulations.\cite{droge09}
Significant anisotropy relative to the Na$^+$ ion distribution is still apparent, however, even at the lower polarizability.
Since the computed probability distribution involves a radial coordinate, we also plot the distribution divided 
by $4\pi R_{\mathrm{cage}}^2$; increased solvation anisotropy then shows up as as a significantly reduced probability at
small radii relative to the sodium case. 
Figure \ref{fig:snap} displays three snapshots of the ion and the first-shell waters in typical
configurations from the default ion polarizability simulations to illustrate the anisotropic solvation. We have
observed some typical configurations of the inner solvation shell, such
as 4, 4+1, or 4+2. A careful investigation of the trajectories suggests
that the first solvation shell of the chloride ion contains many
low energy conformers at room temperature; it is thus difficult to identify a
unique hydration structure. All these conformers exhibit
anisotropic structures, however; similar results have been found in surface{}-like
states reported for Cl$^-$(H$_2$O)$_n$ clusters\cite{tobias1} and Br$^-$ ions in water.\cite{raugei7}  
Wick and Xantheas\cite{wick} have suggested that the extinction of the small cavity on the water-depleted 
side of the ion may create a driving force for anion interfacial activity.  

\subsection{Ion and water dipole moments and charge transfer}

As discussed in a wide range of studies,\cite{mcari97,jungwirth_specific_2006,mmuch05,tchan06,lpere92,dhagb05,wickpol_09} 
anion segregation to the water liquid/vapor interface likely results from a
subtle balance of anion{}-water and water{}-water interactions. In the
following, we study the dipole moments of the ion and the water
molecules in the bulk first solvation shell and charge transfer effects, utilizing high level quantum
chemistry methods. Also we investigate how the polarization of water
molecules in the first solvation shell changes in the presence of the
ion, nearby waters in the first shell, and the surrounding water
molecules beyond the first shell. The goal is to provide an accurate physical
description of the ion{}-water and water{}-water interactions in the
first solvation shell where the classical potential may not suffice for
quantitatively answering polarization questions. 
\ All the calculations are performed at the MP2/aug{}-cc{}-pvdz level based on
the clusters extracted from the trajectories generated by the
simulation with the AMOEBA force field. The effect of the solvating
environment (second and further solvation shells) is handled with
the charge model discussed in Section \ref{sec_Methods}.

Molecular dipole moments are quantities which characterize the charge
distribution in molecules. Charge redistribution or electronic
polarization due to interaction of molecules with the external
environment is directly reflected by the change in the magnitude
of the dipole moments in the various systems we study. Higher-order
polarization effects can be expected to contribute to the interactions, but analysis of the dipoles
can give an initial view of molecular charge redistributions.
We utilize the ChelpG charges and dipoles to investigate the dipole moments
of water molecules and ions averaged over configurations equally
spaced in time by 1 ps. The configurations are taken from the
trajectories of the AMOEBA simulation.  The quantum results are compared to the
dipole moments calculated from the AMOEBA model.

\begin{figure}
\includegraphics[angle=0,scale=1.5]{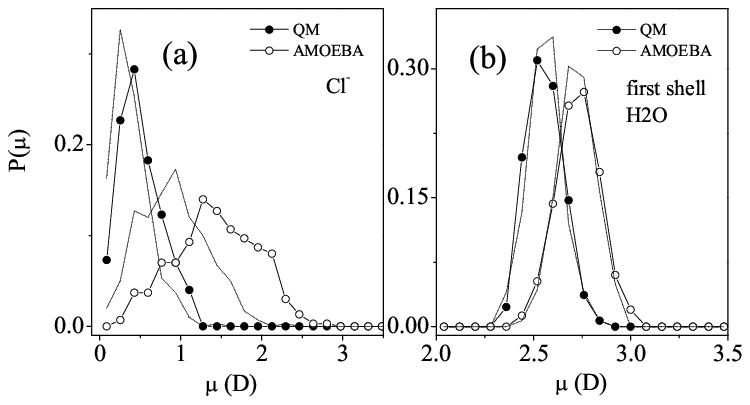}
\caption{The distribution of the magnitude of (a) the anion and (b) 
first-shell water molecule dipole moments (\textit{${\mu}$}). In (a) the solid and 
and filled circles represent the distribution of ${\mu}$ for the Cl$^-$ ion
calculated at the MP2-ChelpG level and the classical AMOEBA model level, respectively
(default chloride ion polarizability case).
The curves without symbols are for the reduced chloride ion polarizability simulations.
In (b) the solid and filled circles represent the distribution of ${\mu}$
for water molecules in the first solvation shell of the anion at the MP2-ChelpG level
and at the classical AMOEBA model level, respectively (default chloride ion polarizability case). 
The curves without symbols are for the 
reduced chloride ion polarizability simulations.} 
\label{fig:dips}
\end{figure}

The calculated anion and first-shell water dipole distributions are shown in Figure \ref{fig:dips}. 
We first present the results for the chloride ion charge distribution.     
As discussed above, the ChelpG fitting was augmented by inclusion of a point dipole on 
the chloride ion to obtain an estimate of the dipole moment from the MP2 calculations.
For the default anion polarizability AMOEBA simulation, the average
dipole moment magnitude of the chloride ion from the MP2-ChelpG method
is 0.6 D
(standard deviation of \ 0.2 D); this differs significantly from
the AMOEBA model result of 1.6 D
(standard deviation of 0.5 D). 
For the reduced anion polarizability AMOEBA simulation, the average
dipole moment magnitudes of the chloride ion from the MP2-ChelpG and AMOEBA methods are
0.5 D (standard deviation of \ 0.2 D) and 1.0 D
 (standard deviation of 0.4 D), respectively. 
Thus the quantum result decreases slightly with reduced anion polarizability, 
presumably due to a less anisotropic hydration environment, 
while the AMOEBA result decreases substantially.
Our MP2 calculations are consistent
with recent extensive AIMD-DFT studies of the chloride ion\cite{masia,eguar09} and the
bromide ion\cite{raugei7} in water, where
average dipole moments of 0.8 D for the chloride ion
and 0.9 D for the bromide ion were observed.  The AIMD simulations utilized a Wannier 
decomposition to estimate the dipole moments; that procedure places full electron charges
on individual atoms. As will be discussed below, we observe a charge transfer of 0.2e from
the chloride ion to the nearby waters.  When that charge transfer effect is 
taken into account, our results are entirely consistent with the estimate from Ref.~\cite{eguar09}
and may provide a more accurate estimate of the average dipole magnitudes. 

On the other hand, Wick and
Xantheas\cite{wick} used the polarizable Dang{}-Chang model to study a chloride ion
in water and found an average ion dipole moment of about 1.3 - 1.5 D, 
which is similar to our AMOEBA calculations (see also Ref.~\cite{droge09}).
\ Thus, the Dang{}-Chang and AMOEBA models appear to significantly overestimate
the anion dipoles. This discrepancy may arise from the
fact that these classical polarizable models contain no or limited damping of the nearby electrostatic 
interactions for the self-consistent polarization calculation.\cite{masia} 
In addition, these classical models 
do not account for charge transfer; force fields to handle charge transfer are 
under active development.\cite{kaminski,chen_qtpie:_2007} 
 Wick\cite{wickpol_09} has recently shown that increased local damping of the charged interactions
during the polarization self-consistency step, leading to smaller anion dipoles, 
reduces the surface affinity of anions at the water liquid-vapor interface.  

In the AMOEBA model, point polarizable dipoles (in addition to the fixed multipoles) are located on
the ion and on each atom within the water molecule. At short
distances there may be diffuse electron distributions, significant electron density overlap, and
possible charge transfer between the ion and the water molecules.  Thus it is a 
challenge for the multipole-based models to reproduce these complex chemical
effects, and the effectiveness of the models should be tested against electronic
structure methods.  
 Work along these lines has been initiated
by Masia and coworkers.\cite{masia2,masia,eguar09}  To further explore these 
issues, we next examine possible 
charge transfer effects between the anions and their neighboring hydration shell.  

\begin{figure}
\includegraphics[angle=0,scale=1.0]{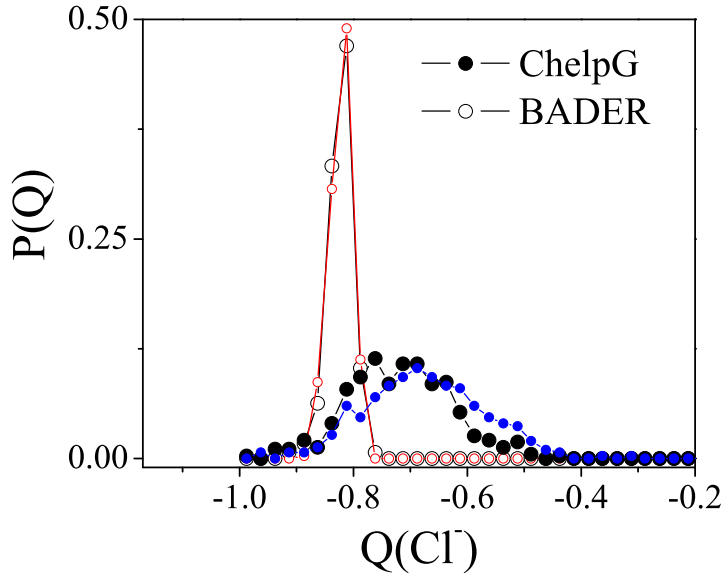}
\caption{Shown are the distributions of charges on the Cl$^-$ ion
calculated using the ChelpG scheme and the QTAIM analysis. 
The curves with filled circles
are from the ChelpG analysis; the black curve is for the default chloride ion polarizability simulation, while
the blue curve is for the reduced polarizability case. The curves labeled with open circles are from 
the QTAIM analysis; the black curve is for the default chloride ion polarizability simulation, while
the red curve is for the reduced polarizability case. The ion/(first-shell water) cluster QM/MM calculations
included electrostatic interactions with more distant waters via the charge model described in Section \ref{sec_Methods}.} 
\label{fig:chgtran}
\end{figure}

Charge transfer has been observed in the study of anion{}-water
clusters and anions in bulk water.\cite{jortner,peraro,bucher,robertson,johnson_arpc,thompson,marenich_polarization_2007} 
To study these effects with our QM/MM model, we
calculated the charge on the chloride ion using both the
ChelpG and the QTAIM methods. Both approaches 
have been shown to be relatively accurate methods for the calculation of the
amount of intermolecular electron transfer.\cite{szefczyk} The calculated
distribution of anion charges is shown in Figure \ref{fig:chgtran}. 
For the default chloride ion polarizability AMOEBA simulation, the average charge of the chloride ion
 is {}-0.7 from the ChelpG method and {}-0.8 from
the QTAIM analysis.  While these effective charge values are relatively close to each other, it can be seen 
from the figure that the distributions differ appreciably; this is not surprising since the charges are
determined by quite different strategies.  Due to its more rigorous physical underpinning,
we view the QTAIM charge calculations as a more accurate 
indication of the net charge on the anions relative to the ChelpG results. 
The calculations suggest charge transfer in the
amount of $\sim 0.2$e from the chloride ion to the nearby
hydration shell.  Modeling the chloride ion with a polarizability of 2 {\AA}$^3$ resulted in only 
slight changes in the distributions of ion charge states (Figure \ref{fig:chgtran}).  
The observation of significant charge transfer is consistent with the previous CPMD study by
Dal Peraro and co{}-workers\cite{peraro} and electronic structure 
calculations on anions in water.\cite{marenich_polarization_2007} 
Rashin {\it et al.}\cite{rashin_charge_2001} applied a different charge partitioning scheme 
and concluded that charge transfer in ion-water clusters is not as significant as the results
discussed above. 

In order to analyze the differences between the ChelpG and QTAIM estimates of charge transfer observed in 
Figure \ref{fig:chgtran}, configurations
were extracted from the AMOEBA simulations for clusters with $N=1-4$ waters.  These random configurations were then
optimized at the MP2 level to generate unique structures for further analysis. The optimized structures were examined here since we are not computing thermal averages over the MD-generated configurations; we simply seek to compare the ChelpG and QTAIM results for some prototype structures of small clusters. The resulting structures are similar to the 1(C$_{\mathrm{s}})$, 2(C$_1$), 3(C$_3$), and 4(C$_4$) structures in Figure 1 of Ref.~\cite{kim_00}. Table \ref{tab:chelpg} lists the computed charges
from the ChelpG method (one charge only on the chloride ion), ChelpG with an added point dipole on the chloride ion, 
and the QTAIM method.  The chloride ion dipoles estimated by the ChelpG calculations are given in the last column.
It is clear from these data that the addition of the dipole in the ChelpG ESP fitting reduces the estimated charge transfer
relative to the fit with only a single charge placed on the ion.  Also, for the $N=3$ and $N=4$ cases, the ChelpG estimate including 
the dipole is close to the QTAIM value.  In principle, higher-order multipoles could also be included in the fitting procedure, but 
we don't follow that direction here.  
Thus it would appear that the ChelpG distribution presented in Figure \ref{fig:chgtran} 
(with only a single charge included on the ion during the ESP fit) 
would likely shift closer to the QTAIM distribution with inclusion of a point dipole (or higher multipoles) in the fitting.  

\begin{figure}
\includegraphics[angle=0,scale=0.7]{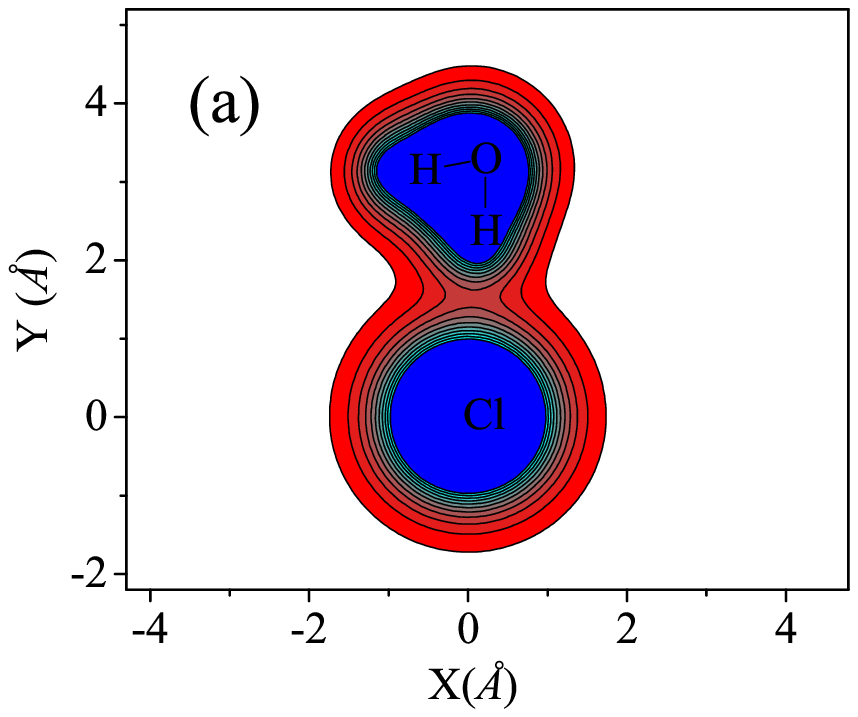} 
\includegraphics[angle=0,scale=0.7]{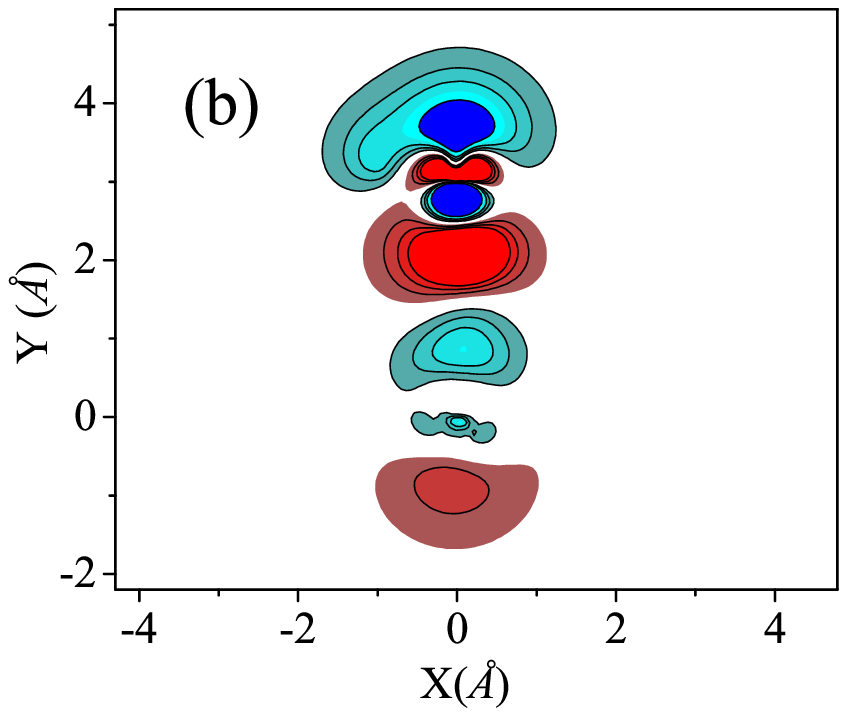}
\caption{(a) Contour plot of the total electron density for the Cl$^-$/(H$_2$O) dimer.  The chosen plane
contains the hydrogen bond between the ion and the water hydrogens. (b) Contour plot of the electron density difference between the Cl$^-$/H$_2$O complex and the separated ion and water molecule. The configuration and plane are the same as in (a). Red indicates reduced electron density, and shades of blue imply increased electron density.} 
\label{fig:den}
\end{figure}

Figure \ref{fig:den} displays an electron density contour plot for the $N=1$ case in the plane determined by the chloride ion
and the water hydrogen and oxygen. Also displayed is the electron density difference contour plot for the same Cl$^-$/H$_2$O complex; the polarization of the chloride ion is apparent from a withdrawal of electron density from the opposite side of the ion from the water molecule and buildup in the direction of the hydrogen bond. The re-distribution of charge on the water molecule is interesting in that there appears to be some depletion in the region of the hydrogen bond to the ion, and a buildup on the other side of the water molecule. These charge redistributions may be related to recent measurements of the oxygen K-edge X-ray absorption spectrum (XAS) of aqueous sodium halide solutions, which suggest perturbations of unoccupied water orbitals due to interactions with anions.\cite{saykally_CT_05}  The electron density and difference contour plots indicate that 
the overall charge distribution is complicated with significant electron density overlap and rearrangement; this suggests a multipole 
expansion to represented the ESP due to the ion may be questionable. On the other hand the QTAIM estimate of 
the ion charge is based on a direct physical criterion derived from the electron density.  
We note that the QTAIM estimate of the charge transfer is roughly additive with increasing 
numbers of waters (after the first water), indicating each additional hydrogen bond makes a similar contribution.  The locations of the hydrogen bonds 
fluctuate widely, however, during thermal sampling, leading to complex and varying charge distributions for the chloride ion.   

\begin{table}
\caption{Charges on the chloride ion estimated by the ChelpG and QTAIM methods. Columns 2-4 display the 
results for the ChelpG method with a single charge on the chloride ion, the ChelpG method with an added 
dipole on the chloride ion, and the QTAIM method, respectively.  The final column is the estimated dipole moment
on the chloride ion obtained from the calculations in column 3.}
\begin{tabular}{ccccc}
$N$ & ChelpG($Q$) & ChelpG-d($Q$)  & QTAIM($Q$) &ChelpG($\mu$-D)\\ 
\hline
1 & -0.938 & -0.977  & -0.918 & 0.24  \\
2 & -0.873 & -0.944  & -0.892 &  0.36 \\
3 & -0.830 & -0.880  & -0.874 &  0.27 \\
4 & -0.805 & -0.830  & -0.847 &  0.10 \\
\hline
\end{tabular}
\label{tab:chelpg}
\end{table}

Next we discuss the results for the dipole distributions in the first-shell water molecules. 
The average value of the dipole moment of the water molecules in the
first solvation shell is close to, but slightly less than, the bulk AMOEBA water value (2.78 D) for both the
MP2-ChelpG and the AMOEBA calculations; the MP2-ChelpG estimate is 2.67 D 
while the AMOEBA model prediction is 2.72 D.  Figure \ref{fig:dips} presents the 
distributions of instantaneous first-shell water dipole magnitudes;
neither the MP2-ChelpG nor the AMOEBA distributions change appreciably with 
reduced chloride ion polarizability in the AMOEBA simulations.
\ Raugei and Klein's\cite{raugei7} CPMD study of the bromide ion in water also
suggested the water molecules in the first shell possess dipole
moments  close to those of bulk water molecules.  Recent extensive AIMD simulations have shown slighly suppressed 
water dipoles in the first solvation shell for the larger halides\cite{eguar09} and the K$^+$ ion,\cite{whitfield_theoretical_2007}
in agreement with our AMOEBA and MP2 results. 

\begin{figure}
\includegraphics[angle=0,scale=1.5]{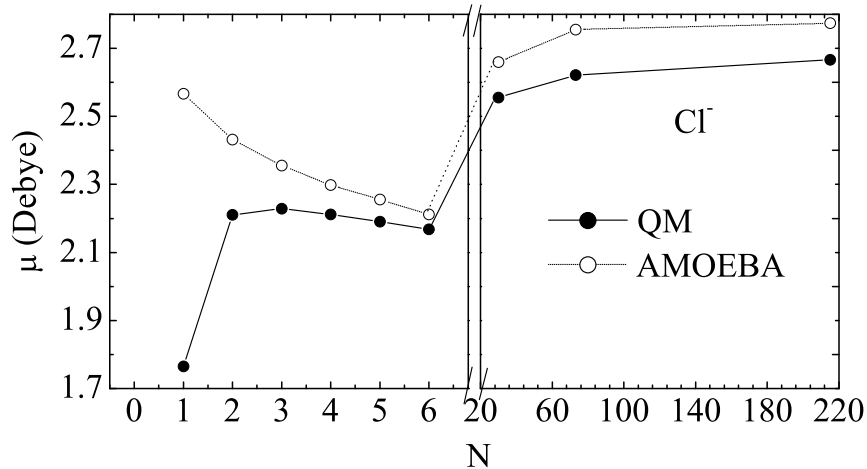}
\caption{The magnitude of the average dipole moment \textit{${\mu}$}
of water molecules neighboring a Cl$^-$ ion as a function of increasing cluster
size \textit{N}. On the left, the system modeled included the ion and the given number 
of water molecules. On the right, 6 waters were treated quantum mechanically in the first shell, 
and the rest of the waters in the QM/MM cluster were represented with AMOEBA charges distributed
as described in Section \ref{sec_Methods}. On the right, average dipoles over the nearest 6 waters are presented, but those waters interact with increasing numbers of surrounding AMOEBA waters. } 
\label{fig:dips_n}
\end{figure}

To further investigate which
interactions are most responsible for the polarization of water
molecules in the first shell and the interplay of ion{}-water and
water{}-water interactions, we studied the magnitude of the average
dipole moment of the water molecules
neighboring the chloride ion as a function of
increasing cluster size \textit{N}, as shown in Figure \ref{fig:dips_n}. This was done to account for the impact of increasing numbers of waters surrounding the ion. Only the default chloride
ion polarizability case was considered in examining the size dependence, since the anion 
polarizability had a relatively small effect on the water dipole distributions.  The first 6 water molecules
were treated at the MP2 level, while more distant waters were handled with the effective
charge model obtained from the AMOEBA simulation. 

For this analysis, we present the thermally averaged results for clusters of the ion
with the $N$ nearest waters extracted from the simulations and plot the results as 
a function of $N$. On the left side of Figure \ref{fig:dips_n}, average dipole magnitudes were
computed for the $N$ nearest waters at the quantum level with no interactions with more distant waters.
As we increase the ion{}-water cluster size ($N=1-6$), we find that
the average dipole moment of the water molecules in the first hydration
shell of the ion (at the QM/MM level) first increases to a value around 2.2 D and then stays
roughly constant (decreasing slightly with increasing $N$). 
We note that, in our calculations, the numbering is such that the second and successive water molecules are chosen as the nearest water to the preceding water oxygen.  This choice was made in order to explore
the effects of nearby water-water interactions.
The average distance between the two water molecules, measured by the
distance between the two O atoms, is around 3.5 {\AA}. \ The addition
of a second water molecule significantly increases the polarization of the
first water molecule, even though there is no direct hydrogen bonding
between the two water molecules. The appreciable deviation between the quantum and AMOEBA results for the
$N=1$ case may be tied to the larger charge transfer to the closest water discussed above, and may 
reflect limitations of the classical polarizable model. 
Further addition of water molecules in the
first hydration shell doesn't substantially affect the polarization of water
molecules on average. Previous reports examining polarization for
hydration shell water molecules of anions displayed a different $N$ dependence 
for the water dipoles.\cite{krekeler3} 
This is likely because of the different 
Cl$^-$/(H$_2$O)$_n$ cluster structures investigated. We utilize the instantaneous
cluster structures extracted from the molecular dynamics simulation, 
while Ref.~\cite{krekeler3} used optimized gas phase
cluster structures. Comparing our MP2 calculations with the AMOEBA
calculations, we find the AMOEBA model slightly overestimates the dipole moments
of the water molecules in the first shell for $N\ge2$, with better agreement as $N$ increases.

Finally, we study the effect of the second solvation shell and further shell
water molecules on the polarization of the first shell water molecules.
On the right of Figure \ref{fig:dips_n}, the average dipole magnitudes of the nearest 6 waters are presented, but those waters 
interact electrostatically with the remainder of the $N-6$ waters handled at the AMOEBA level. 
We find that the addition of the second solvation shell (represented as sets of charges so as to 
reproduce the AMOEBA multipole potential) with formation
of direct hydrogen bonding to those external waters increases the
average dipole moment of the first shell water molecules further by about
20\%. Third and further solvation shells don't appear to have a
large impact on the polarization of the first shell water molecules.
As discussed above, upon full hydration the average dipole magnitude is 2.67 D for the
inner-shell water molecules, which is slighly smaller than the AMOEBA value of 2.72 D and close to the AMOEBA value for bulk water
(2.78 D). We note that the water dipole values presented here are somewhat below those observed in AIMD-DFT 
simulations;\cite{raugei7,heuft_density_2003,eguar09} an experimental estimate for bulk water is 2.9 D.\cite{soper00}  
The results support the picture suggested by Raugei
and Klein,\cite{raugei7} Krekeler and Delle Site,\cite{krekeler2} and 
Gu\'{a}rdia, Skarmoutsos, and Masia\cite{eguar09} that the anion scarcely influences the
polarization of the water molecules in the first solvation shell; in fact the inner-shell waters appear to have average dipole magnitudes slightly reduced from the bulk value.  We
observe that the first-shell water{}-water interactions (modified by interactions with the ion) and the
H{}-bonding environment from the second solvating shell are the major
factors that determine the polarization of the first shell water
molecules. This result supports the idea suggested by Devereux and
Popelier that the local H{}-bonding environment of a water molecule
determines its charge distribution and electronic properties.\cite{devereux}

\section{summary and discussion}
\label{sec_discussion}


This paper has examined the local structure and polarization involved in chloride ion hydration with a QM/MM treatment.  Classical polarizable force field molecular dynamics simulations were performed with the AMOEBA model. Clusters were extracted from the simulations for further analysis with quantum chemical methods at the MP2-level of electronic structure theory. The clusters treated quantum mechanically involved the ion and the nearest waters in the first solvation shell. The inner-shell clusters interacted with more distant waters modeled with AMOEBA charge distributions.  In particular, we examined the average dipole magnitudes for the chloride ion and the nearby waters, and charge transfer from the anion to the waters. The role of interactions with waters outside the first shell was explored by successively increasing the size of the QM/MM clusters. 

In the classical polarizable AMOEBA simulations, anisotropic solvation structures were observed around the chloride ion, consistent with other classical models\cite{wick} and AIMD simulations.\cite{raugei7} The anisotropy decreases with decreasing chloride ion polarizability, but is still significant with the polarizability reduced by a factor of 2.  The solvation anisotropy is clear from visual inspection of representative configurations, and from examination of the location of the center-of-mass of the inner-shell waters relative to the chloride ion location.  The most probable coordination number observed in the AMOEBA simulations is 6 based on a standard hydrogen-bonding criterion.

The first-shell water molecule dipoles computed quantum mechanically (in the external field of surrounding waters) are slightly reduced relative to the AMOEBA model, and both exhibit slightly reduced values compared with bulk AMOEBA water.   This result is somewhat surprising, but is consistent with recent AIMD-DFT simulations.\cite{whitfield_theoretical_2007,eguar09}  Reducing the chloride ion polarizability in the AMOEBA simulations had little effect on the first-shell water dipoles.  We find that water-water interactions within the inner shell and further interactions with more distant waters are all crucial in determining the polarization state of the inner-shell waters. In fact, the water-water interactions appear to dominate over the ion-water interactions in the polarization.\cite{krekeler2}  

The average chloride ion dipole moments computed quantum mechanically are significantly lower than those observed in the AMOEBA simulations, however. Such behavior has also been found in several AIMD simulations,\cite{raugei7,heuft_density_2003,eguar09} and it thus appears that the AMOEBA model, even with Thole-type damping incorporated for the nearby electrostatic interactions, is over-polarizing the chloride ion.  Since we showed in another study\cite{droge09} that the polarization is linked to solvation anisotropy, this finding raises the issue that some of the previous classical polarizable models may be consistently over-estimating solvation anisotropy.  The quantum mechanical ion dipoles were reduced slightly when the configurations were sampled with the lower chloride polarizability, presumably due to the lower anisotropy in the first-shell waters.  The issue of designing damping functions that properly represent the ion dipoles in classical simulations has been addressed by Masia.\cite{masia} 

In addition, we observe significant charge transfer from the chloride ion to the nearby waters (0.2e from the QTAIM method).  Charge transfer has been implicated in a wide range of biophysical solvation problems\cite{collins_ions_2007} and has been previously observed in DFT simulations\cite{peraro} and quantum chemical studies.\cite{marenich_polarization_2007,thompson_frequency_2000,srame08} It is possible that the over-polarized ions observed in the AMOEBA model somehow mimic this charge transfer effect in an average way, but further work is necessary to quantify the relation of the classical models to the more realistic quantum calculations (the results for the small $N$ cases in Figure \ref{fig:dips_n} suggest that some of the essential physics is missing from the multipole representation, however).  Since most models used to study the surface affinity of anions to the water liquid-vapor interface have employed no electrostatic damping, or limited versions as in the AMOEBA model, these issues should be revisited in modeling the specific ion effects. A recent study has confirmed that reduced ion polarization in turn reduces anion surface affinity.\cite{wickpol_09}

We have chosen a QM/MM approach with an MP2-level treatment of the inner shell for several reasons.  
First, this level of theory should generate relatively accurate charge distributions for analysis of local polarization around the ions.\cite{kim_00}  Second, MP2 calculations  include many-body dispersion interactions at a decent level of accuracy, and we believe this may be important for modeling ion-water and water-water
interactions accurately.\cite{wkunz04,santra_accuracy_2008}  
Third, we plan to couple this QM/MM approach to computations of solvation free energies via quasi-chemical theory.\cite{ourbook,droge08,droge09}  
In the quasi-chemical approach, the solvation free energy is exactly partitioned into inner-shell, outer-shell packing, and outer-shell long-range contributions. The conditioning inherent in this partitioning allows for a mean-field treatment of the long-range contribution. We have observed in another study\cite{droge09} that relatively small conditioning radii already lead to Gaussian behavior in the long-ranged contribution.  With such small conditioning radii, most of the hydration free energy is contained in the long-ranged part, and that term can then be examined for the various contributions to ion specificity.  We plan to compute the solvation free energies in our MP2-level QM/MM model, and further explore the various contributions to the free energy (first-order electrostatic, induction, dispersion, charge transfer) via an 
SAPT energy division.\cite{amisq05,bukowski_predictions_2007}  
Even though there have been multiple indications that anion polarization is crucial in hydration, it may be possible that, due to the observed charge transfer in quantum models and over-polarization in the classical models, that higher-level quantum mechanical treatments are necessary to accurately represent the local hydration structure and energetics.

Computing effective water and ion charges and dipoles can provide helpful insights into the local solvation structure.  
These charges and multipoles could then serve as input for the development of more 
refined molecular dynamics force fields.  An alternative suggestion, however, is that 
the ion solvation environment involves complicated and diffuse charge distributions best represented with electronic structure methods.   The 
quasi-chemical treatment of solvation thermodynamics provides a natural statistical mechanical framework for representing the first solvation
shell at an accurate quantum mechanical level, with a lower-level representation of more distant 
waters.\cite{asthagiri_quasi-chemical_2003,asthagiri_absolute_2003,varma_structural_2008}

\section{Acknowledgments}
We gratefully acknowledge the support of NSF grant CHE-0709560 and the DOE Computational Science Graduate Fellowship (DE-FG02-97ER25308, to DMR) for the support of this work. We acknowledge the Ohio Supercomputer Center for a grant of computing time. We thank Marco Masia for comments on the paper.


\end{document}